\documentclass{article}

\usepackage{amsmath}
\usepackage{graphicx}
\usepackage{dcolumn}
\usepackage{bm}
\usepackage{hyperref}

\begin{document}

\title{Maximum Entropy and the Variational Method in Statistical Mechanics:\\
an Application to Simple Fluids }
\author{Chih-Yuan Tseng\thanks{%
Present address: Department of Physics, National Central University, Jhongli
320, Taiwan ROC; TEL.: 886-3-4227151-65368; FAX: 886-3-4251175; E-mail:
richard@pooh.phy.ncu.edu.tw} and Ariel Caticha\thanks{%
E-mail: ariel@albany.edu} \\
Department of Physics, University at Albany-SUNY\\
Albany, NY 12222 USA}
\date{Received: date / Accepted: date}
\maketitle

\begin{abstract}
We develop the method of Maximum Entropy (ME) as a technique to generate
approximations to probability distributions. The central results consist in
(a) justifying the use of relative entropy as the uniquely natural criterion
to select a ``best'' approximation from within a family of trial
distributions, and (b) to quantify the extent to which non-optimal trial
distributions are ruled out. The Bogoliubov variational method is shown to
be included as a special case. As an illustration we apply our method to
simple fluids. In a first use of the ME method the ``exact'' canonical
distribution is approximated by that of a fluid of hard spheres and ME is
used to select the optimal value of the hard-sphere diameter. A second, more
refined application of the ME method approximates the ``exact'' distribution
by a suitably weighed average over different hard-sphere diameters and leads
to a considerable improvement in accounting for the soft-core nature of the
interatomic potential. As a specific example, the radial distribution
function and the equation of state for a Lennard-Jones fluid (Argon) are
compared with results from molecular dynamics simulations.\newline
\textbf{KEY WORDS}: Variational method; maximum entropy; simple fluids 
\end{abstract}


\section{Introduction}

\label{intro} A common problem is that our beliefs and our knowledge are
often represented by a probability distribution $P(x)$ that is too
complicated to be useful for practical calculations and we need to find a
more tractable approximation. A possible solution is to identify a family of
trial distributions $\{p(x)\}$ and select the member of the family that is
closest\ to the exact\ distribution $P(x)$. The problem, of course, is that
it is not clear what one means by `closest'. We could minimize 
\begin{equation}
\int dx\,\left[ p(x)-P(x)\right] ^{2}~,
\end{equation}%
but why this particular functional and not another? And also, why limit
oneself to an approximation by a single member of the trial family? Why not
consider some linear combination of the trial distributions, some kind of
average over the trial family? But then, how should we choose the optimal
weight assigned to each $p(x)$? And again, what does `optimal' mean? We
propose to tackle these questions using the method of maximum entropy.

The method of Maximum Entropy as advocated by Jaynes (MaxEnt) is a method to
assign probabilities on the basis of partial information \cite{Jaynes57}.
The core of the MaxEnt method resides in interpreting entropy, through the
Shannon axioms, as a measure of the ``amount of uncertainty'' or of the
``amount of information that is missing'' in a probability distribution.
This was an important step forward because it extended the applicability of
the notion of entropy far beyond its original roots in thermodynamics but it
was not entirely free from objections. For example, one problem is that the
Shannon axioms refer to probabilities of discrete variables; the entropy of
continuous distributions was not defined. A second, perhaps more serious
problem is that other measures of uncertainty were subsequently found. This
immediately raised the question of which among the alternatives should one
choose and motivated attempts to justify the method of maximum entropy
directly as a method of inductive inference without referring to any
specific measure of uncertainty \cite{ShoreJohnson80}-\cite{Caticha03}. What
eventually emerged is an extended form of the method of Maximum Entropy,
which we abbreviate as ME to distinguish it from the original MaxEnt. The
core of ME is a concept of (relative) entropy that reduces to the usual
entropy in the special case of a uniform prior or uniform background measure
and that does not require an interpretation in terms of heat, disorder,
uncertainty, or even in terms of an amount of information: in the context of
ME \emph{entropy needs no interpretation}.

The ME method is designed for processing information, for updating from a
prior probability distribution (which in statistical mechanics is generally
postulated to be uniform) to a posterior distribution when new information
in the form of constraints becomes available \cite{footnote1}. The chosen
posterior distribution should coincide with the prior as closely as
possible: ME implements the minimal changes that are required to satisfy the
new constraints. The significance of this development is that it allows one
to tackle problems lying beyond the restricted scope of MaxEnt \cite%
{Caticha03b}. 

The main purpose of this paper is to extend the use of the ME method beyond
processing information to a method to generate approximations \cite%
{footnote2}; a second purpose is to illustrate the method by applying it to
simple fluids.

The general formalism, which is the main result of this paper, is developed
in section ~\ref{sec2}. In section ~\ref{sec2-1} we justify the use of
relative entropy as the uniquely natural criterion to select the best
approximation. In section ~\ref{sec2-2}, as a simple illustration, the ME
method is applied to trial distributions of the canonical Boltzmann-Gibbs
form. The result, when expressed in terms of free energies, is recognized as
the Bogoliubov variational principle \cite{Callen85}, which is now seen as a
special case of ME \cite{Opper01}. The first part of our paper concludes in
section ~\ref{sec2-3} with a quantitative analysis of the extent to which
the distribution that maximizes the relative entropy is preferred over other
members of the trial family. We show that our entropic argument does not
completely rule out those distributions that fail to maximize the entropy.
Therefore a suitably weighed average over the whole family of trial
distributions with the optimal weight calculated using the ME method
provides a better approximation than can be expected from any individual
member of the trial family.

The discussion up to this point is general -- the ME method of generating
best approximations is of general applicability -- but it is useful to see
how it is applied to a specific problem. In the second part of the paper we
apply the ME method to simple classical fluids. The study of classical
fluids is an old and mature field. There exist extensive treatments in many
excellent books and reviews \cite{BarkerHenderson76}-\cite{Kalikmanov02}.
Our goal is to use this well-explored but still non-trivial field as a
testing ground for the ME method. At this early stage in the development of
the ME method we are not concerned with contributing to the study of the
fluids themselves. Indeed, the reason we choose such a well-explored
problem, is not because we are eager to find new results about fluids but
because we can compare our results with those obtained using alternative
approximation methods already in existence. We find that our results are
quite competitive with those obtained using the best perturbative methods
developed to date.

The practical success or failure of the method hinges on choosing a family
of trial distributions that incorporates the information that is relevant to
the problem of interest. In our example -- simple fluids -- it turns out
that their structure is dominated, particularly at liquid densities, by the
short-range repulsions between molecules. The effects of the long-range
attractions are averaged over many molecules and are less important, except
at low densities. Accordingly we choose a fluid of hard spheres \cite%
{BarkerHenderson76}-\cite{Kalikmanov02} as our (almost) tractable trial
model (section ~\ref{sec3}). The ME method is then used (section ~\ref{sec4}%
) to select the optimal value of the hard-sphere diameter. This is
equivalent to applying the Bogoliubov variational principle and reproduces
the results obtained by by Mansoori et al. \cite{Mansoori69}, and
independently by Stell at al. \cite{Stell70}.

A very different approximation scheme can be traced back to Lewis who used a
maximum entropy argument to derive the Boltzmann equation \cite{Lewis67},
and was further developed by Karkheck, Stell and coworkers \cite{Karkheck82}%
. Instead of selecting the optimal distribution from within a family of
trial approximations, which is what we do here, in their \textquotedblleft
kinetic variational theory\textquotedblright\ an entropy is maximized to
determine the optimal closure for the BBGKY hierarchy of equations. Their
method is suitable for tackling transport problems in dense simple fluids
with potentials that are more realistic than just hard spheres. 

Alternative approaches to the study of fluids include perturbation theory.
In such schemes the intermolecular potential $u$ is written as $u_{0}+\delta
u$, where $u_{0}$ is a strong short-range repulsion, which is eventually
approximated by a hard-sphere potential $u_{hs}$, and $\delta u$ is a
long-range attraction treated as a perturbation. At high densities first
order theory is quite accurate but at lower densities higher-order
corrections must be included. There exist several proposals for how to
separate $u$ into $u_{0}$ and $\delta u$, for choosing the particular stage
in the calculation where $u_{0}$ is replaced by $u_{hs}$, and also for
choosing the best hard-sphere diameter. The most successful are those by
Barker and Henderson \cite{BarkerHenderson76} and by Weeks, Chandler and
Andersen (WCA) \cite{WCA}. The latter succeeds in using the hard-sphere
potential $u_{hs}$ while effectively representing some of the effects of the
soft-core potential $u_{0}$. For a recent discussion of some of the
strengths and limitations of the perturbative approach see \cite{Germain02}.

An advantage of the variational and the ME methods over the perturbative
approaches is that there is no need for ad hoc criteria dictating how to
separate $u$ into $u_{0}$ and $\delta u$, and how to choose a hard-sphere
diameter. A disadvantage of the standard variational approach is that it
fails to take the softness of the repulsive core into account, which leads,
at high temperatures, to results that are inferior to the perturbative
approaches. As we shall see below this limitation does not apply to the ME
method.

The standard variational approach allows one to select the optimal value of
a parameter, in this case the hard-sphere diameter; all non-optimal values
are ruled out. But, as discussed in \cite{Caticha03} and \cite{Caticha00},
the ME method allows one to quantify the extent to which non-optimal values
should contribute. In this more complete use of the ME method (presented in
section ~\ref{sec5}) the ``exact'' probability distribution of the fluid is
approximated by a statistical mixture of distributions corresponding to hard
spheres of different diameters. This is a rather simple and elegant way to
take into account the fact that the actual atoms are not hard spheres. The
full ME analysis leads to significant improvements over the variational
method.

In section ~\ref{sec6} we test our method by comparing its predictions for a
Lennard-Jones model for Argon with the numerical molecular dynamics
simulation data (\cite{Verlet68}, \cite{HansenVerlet69}). We find that the
ME predictions for thermodynamic variables and for the radial distribution
function are considerable improvements over the standard Bogoliubov
variational result, and are comparable to the perturbative results. Finally,
our conclusions and some remarks on further improvements are presented in
section ~\ref{sec7}.

\section{Using ME to generate approximations}

\label{sec2} The goal is to select from a family of trial distributions $%
p(x) $ (the possible posteriors) that which is closest to a given ``exact''
distribution $P(x)$ (the prior). The family of trials can be defined in a
variety of ways. For example, one can specify a functional form $p_{\theta
}(x)$, each member of the trial family being labelled by one or more
parameters $\theta $. More generally, one could define the trial family in a
non-parametric way by specifying various constraints.

The discussion below which develops the use of ME to generate approximations
follows \cite{Caticha03} where the ME\ method is developed for a different
purpose, namely, as a method for processing information to update
distributions.

\subsection{The logic behind the ME method}

\label{sec2-1} The selection of one distribution from within a family is
achieved by ranking the distributions according to increasing \emph{%
preference}. Before we address the issue of what it is that makes one
distribution preferable over another we note that there is one feature we
must impose on any ranking scheme. The feature is transitivity: if
distribution $p_{1}$ is preferred over distribution $p_{2}$, and $p_{2}$ is
preferred over $p_{3}$, then $p_{1} $ is preferred over $p_{3}$. Such
transitive rankings are implemented by assigning to each $p(x)$ a real
number $S[p]$ which we call the ``entropy'' of $p$. The numbers $S[p]$ are
such that if $p_{1}$ is preferred over $p_{2}$, then $S[p_{1}]>S[p_{2}]$.
Thus, by design, the ``best'' approximation $p$ is that which maximizes the
entropy $S[p]$.

Next we determine the functional form of $S[p]$. This is the general rule
that provides the criterion for preference; in our case it defines what we
mean by the ``closest'' or ``best'' approximation. The basic strategy \cite%
{Skilling88} is one of induction: (1) if a general rule exists, then it must
apply to special cases; (2) if in a certain special case we know which is
the best approximation, then this knowledge can be used to constrain the
form of $S[p]$; and finally, (3) if enough special cases are known, then $%
S[p]$ will be completely determined.

The known special cases are called the \textquotedblleft
axioms\textquotedblright\ of ME and they reflect the conviction that one
should not change one's mind frivolously, that whatever information was
originally codified into the exact $P(x)$ is important and should be
preserved. The selected trial distribution should coincide with the exact
one as closely as possible and one should only tolerate those minimal
changes that are demanded by the information that defines the family of
trials. Three axioms and their consequences are listed below. Detailed
proofs and more extensive comments are given in \cite{Caticha03}.

\textbf{Axiom 1: Locality}. \emph{Local information has local effects.} If
the constraints that define the trial family do \emph{not} refer to a
certain domain $D$ of the variable $x$, then the conditional probabilities $%
p(x|D)$ need \emph{not} be revised, $p(x|D)=P(x|D)$. The consequence of the
axiom is that non-overlapping domains of $x$ contribute additively to the
entropy: $S[p]=\int dx\,F(p(x),x)$ where $F$ is some unknown function.

The motivation behind this axiom is the following. Suppose that the
information, that is, the constraint, that defines the trial family, does
not refer to a particular subdomain $D$ of the variable $x$. This means that
the probability of $x$ conditional on its being in $D$ is completely
unsconstrained, and $p(x|D)$ can take whatever value we desire. In other
words, the family of trial distributions includes members that are capable
of reproducing the conditional probability $P(x|D)$ exactly. Axiom 1 says
that if we can reproduce $P(x|D)$ exactly then we should; that is, among the
possible trials choose one such that $p(x|D)=P(x|D)$. Clearly this is not
just a good approximation, it is the best we can possibly do.

\textbf{Axiom 2: Coordinate invariance.} \emph{The ranking should not depend
on the system of coordinates. }The coordinates that label the points $x$ are
arbitrary; they carry no information. The consequence of this axiom is that $%
S[p]=\int dx\,p(x)f(p(x)/m(x))$ involves coordinate invariants such as $%
dx\,p(x)$ and $p(x)/m(x)$, where the function $m(x)$ is a density, and both
functions $m$ and $f$ are, at this point, unknown.

Next we make a second use of Axiom 1 (locality). When there are no
constraints at all the family of trials includes the exact $P(x)$ and the
selected trial should coincide with $P(x)$; that is, the best approximation
to $P(x)$ is $P(x)$ itself. The consequence is that up to normalization the
previously unknown density $m(x)$ is the exact distribution $P(x)$.

\textbf{Axiom 3:\ Consistency for independent subsystems}. \emph{When a
system is composed of subsystems that are believed to be independent it
should not matter whether the approximation procedure treats them separately
or jointly.} Specifically, if $x=(x_{1},x_{2})$, and the exact distributions
for the subsystems, $P_{1}(x_{1})$ and $P_{2}(x_{2})$, are respectively
approximated by $p_{1}(x_{1})$ and $p_{2}(x_{2})$, then the exact
distribution for the whole system $P_{1}(x_{1})P_{2}(x_{2})$ should be
approximated by $p_{1}(x_{1})p_{2}(x_{2})$. This axiom restricts the
function $f$ to be a logarithm.

The overall consequence of these axioms is that the trial approximations $%
p(x)$ should be ranked relative to the exact $P(x)$ according to their
(relative) entropy, 
\begin{equation}
S[p|P]=-\int dx\,p(x)\log \frac{p(x)}{P(x)}.  \label{S[p|P]}
\end{equation}%
The derivation has singled out the relative entropy $S[p|P]$ as \emph{the
unique functional to be used for the purpose of selecting an optimal
approximation}. Other functionals, may be useful for other purposes, but
they are not a generalization from the simple cases described in the axioms
above.

\subsection{A special case: the variational method}

\label{sec2-2} As an illustration consider a system with microstates
labelled by $q$ (for example, the location in phase space or perhaps the
values of spin variables). Let the probability that the microstate lies
within a particular range $dq$ be given by the canonical distribution 
\begin{equation}
P(q)dq=\frac{e^{-\beta H(q)}}{Z}dq,  \label{P(q)}
\end{equation}%
where 
\begin{equation}
Z=e^{-\beta F}=\int dq\,e^{-\beta H(q)}.
\end{equation}%
We want to approximate the ``exact'' $P$ by a more tractable distribution $p$%
. The first step is to identify a family of trial distributions that are
simple enough that actual calculations are feasible and that incorporates
the appropriate relevant information. This step is difficult because there
is no known systematic procedure to carry it out; it is a matter of trial
and error guided by intuition. We will consider a family of trial
distributions $p_{\theta }(q)$ that are canonical with a Hamiltonian $%
H_{\theta }(q)$ that depends on parameters $\theta =\{\theta ^{1},\theta
^{2},\ldots ,\theta ^{n}\}$, 
\begin{equation}
p_{\theta }(q)dq=\frac{e^{-\beta H_{\theta }(q)}}{Z_{\theta }}dq,
\label{ptheta(q)}
\end{equation}%
where 
\begin{equation}
Z_{\theta }=e^{-\beta F_{\theta }}=\int dq\,e^{-\beta H_{\theta }(q)}.
\end{equation}%
The next step is to select the $p_{\theta }$ that maximizes $S[p_{\theta
}|P] $. Substituting into eq.(\ref{S[p|P]}) gives, 
\begin{equation}
S[p_{\theta }|P]=\beta \left( \langle H_{\theta }-H\rangle _{\theta
}-F_{\theta }+F\right) \,,  \label{S[ptheta|P]}
\end{equation}%
where $\langle \ldots \rangle _{\theta }$ refers to averages over the trial $%
p_{\theta }$. The inequality $S[p_{\theta }|P]\leq 0$, can then be written
as 
\begin{equation}
F\leq F_{\theta }+\langle H-H_{\theta }\rangle _{\theta }\,.
\end{equation}%
Thus, maximizing $S[p_{\theta }|P]$ is equivalent to minimizing the quantity 
$F_{\theta }+\langle H-H_{\theta }\rangle _{\theta }$. This alternate form
of the variational principle and its use to generate approximations is well
known. It is usually associated with the name of Bogoliubov \cite{Callen85}
and it is the main technique to generate mean field approximations for
discrete systems of spins on a lattice. What is perhaps not as widely known
is that the Bogoliubov variational principle is just the special case of
applying the ME method to trial distributions of the canonical
Boltzmann-Gibbs form.

\subsection{To what extent are non-optimal distributions ruled out?}

\label{sec2-3} The example above does not exhaust the power of the ME\
method: we have found a way to determine the optimal choice of $\theta $ but
ME allows us to go further and quantify the extent to which the optimal $%
\theta $ is preferred over other non-optimal values (\cite{Caticha03}, \cite%
{Caticha00}).

To what extent do we believe that any particular $\theta $ should have been
chosen? This is a question about the probability of $\theta $, $p(\theta )$.
The original problem of assigning a probability to $q$ is now broadened into
assigning probabilities to $q$ and $\theta $.\ Here, we use ME not just to
find the optimal approximation $p(q)$ but also to find the optimal joint
distribution $p(q,\theta )$.

Notice that this is the kind of problem where it is necessary to adopt a
Bayesian interpretation of probabilities. Within a frequentist
interpretation it makes no sense to talk about $p(\theta )$ or $p(q,\theta )$
because $\theta $ is not a random variable; the value of $\theta $ is
unknown but it is not random.

To proceed we must address two questions. First, what is the prior
distribution, that is, what do we know about $q$ and $\theta $ before we are
given the constraints? And second, what are the constraints that define the
family of joint trials $p(q,\theta )$?

When we know nothing about the $\theta $s we know neither their physical
meaning nor whether there is any relation to the $q$. A joint prior $%
m(q,\theta )$ that reflects this lack of correlations is a product, $%
m(q,\theta )=P(q)\mu (\theta )$, where $P(q)$ is the ``exact'' distribution,
say eq.(\ref{P(q)}), and $\mu (\theta )$\ should reflect our ignorance about 
$\theta $: it should be as uniform as possible and make every volume element
in $\theta $ space as likely as any other. But if we know absolutely nothing
about $\theta $ we also do not know how to measure volumes in $\theta $
space.

To make further progress we need the additional information that provides
meaning to the $\theta $s, namely, that they are parameters labeling the
family of distributions $p_{\theta }(q)$. Remarkably this is sufficient to
allow us to introduce a measure of distance in $\theta $ space that is both
natural and unique: we define the distance between $\theta $ and $\theta
+d\theta $ to be the same as the distance between the corresponding $%
p_{\theta }$ and $p_{\theta +d\theta }$ which is given by the Fisher-Rao
metric $d\ell ^{2}=\gamma _{ij}d\theta ^{i}d\theta ^{j}$, where 
\begin{equation}
\gamma _{ij}=\int dq\,p_{\theta }(q)\frac{\partial \log ~p_{\theta }(q)}{%
\partial \theta ^{i}}\frac{\partial \log ~p_{\theta }(q)}{\partial \theta
^{j}}.  \label{Fisher metric}
\end{equation}%
This is the only Riemannian metric that takes proper account of the fact
that the $\theta $s are not just structureless points but represent
probability distributions \cite{Cencov81}. Accordingly, the volume of a
small region $d\theta $ is $\gamma ^{1/2}(\theta )d\theta $, where $\gamma
(\theta )$ is the determinant of $\gamma _{ij}$. Up to an irrelevant
normalization, the distribution $\mu (\theta )$ that is uniform in $\theta $
is given by $\mu (\theta )=\gamma ^{1/2}(\theta )$.

The second question about the constraints that define the family of trials
is straightforward: of all joint distributions $p(q,\theta )=p(\theta
)p(q|\theta )$ we are only interested in the subset of those distributions
such that $p(q|\theta )=p_{\theta }(q)$. Therefore, the trials $p(q,\theta )$
are constrained to be of the form $p(q,\theta )=p(\theta )p_{\theta }(q)$.

Now we allow the ME method to take over: the best approximation $p(q,\theta
) $ to the joint distribution $P(q)\gamma ^{1/2}(\theta )$ is obtained by
maximizing the entropy%
\begin{equation}
\sigma \lbrack p(q,\theta )|\gamma ^{1/2}P]=-\int dq\,d\theta \,p(\theta
)p_{\theta }(q)\,\log \frac{p(\theta )p_{\theta }(q)}{\gamma ^{1/2}(\theta
)P(q)},  \label{sigma[p]}
\end{equation}%
by varying $p(\theta )$ subject to $\int d\theta \,p(\theta )=1$. The final
result for the probability that $\theta $ lies within the small volume $%
\gamma ^{1/2}(\theta )d\theta $ is 
\begin{equation}
p(\theta )d\theta =\frac{1}{\zeta }\,\,e^{S[p_{\theta }|P]}\gamma
^{1/2}(\theta )d\theta \text{~},  \label{p(theta)}
\end{equation}%
where $\zeta $ is a normalization constant.

Eq.(\ref{p(theta)}) tells us that the preferred value of $\theta $ maximizes
the entropy $S[p_{\theta }|P]$, Eq.(\ref{S[ptheta|P]}), because this
maximizes the probability density $\exp S[p_{\theta }|P]$. It also tells us
the degree to which values of $\theta $ away from the maximum are ruled out.
For macroscopic systems the preference for the ME distribution can be
overwhelming. Note also that the density $\exp S[p_{\theta }|P]$ is a scalar
function and the presence of the Jacobian factor $\gamma ^{1/2}(\theta )$
makes Eq.(\ref{p(theta)}) manifestly invariant under changes of the
coordinates $\theta $.

Finally, now that we have determined the joint distribution $p(q,\theta
)=p(\theta )p_{\theta }(q)$ we can marginalize $\theta $ and use the average 
\begin{equation}
\bar{p}(q)=\int d\theta \,p(\theta )p_{\theta }(q)  \label{pbar}
\end{equation}%
as the best approximation we can construct out of the given trial family.
This approximation is expected to be better than any individual $p_{\theta
}(q)$ for the same reason that the mean is expected to be a better estimator
than the mode -- it minimizes the variance. 

This idea of introducing mixtures of probability distributions as in eq.(\ref%
{pbar}) might seem strange at first sight but it is actually quite natural.
For example, if we know that a system is in thermal equilibrium at
temperature $T$ then we describe its macrostate using the canonical
distribution. But if we are uncertain about the actual value of the
temperature then a better description is given by a suitable weighted
average over $T$. Eq.(\ref{p(theta)}) gives the appropriate weights.

The procedure above is mathematically straightforward but the shift from the
original problem of assigning a probability to $q$ into the new problem of
assigning probabilities to $q$ and $\theta $ can be a potential source of
confusion. It might appear that the maximization of the two entropies $S$ in
eq.(\ref{S[p|P]}) and $\sigma $ in eq.(\ref{sigma[p]}) has lead to two
different best approximations: one is $p_{\theta }(q)$ with $\theta $
maximizing eq.(\ref{S[ptheta|P]}), and the other is $\bar{p}(q)$ in eq.(\ref%
{pbar}). How can we have two different answers to the same question? The
answer is that we actually have two different questions. Maximizing $S$
answers the question: \textquotedblleft What is the best single $p_{\theta
}(q)$? Or, what is the best approximation obtainable in terms of a single
trial?\textquotedblright\ Maximizing $\sigma $ answers a different question:
\textquotedblleft What is the best joint distribution $p(q,\theta )$? Or,
equivalently, what is the best approximation when we are not restricted to a
single trial?\textquotedblright\ 

This concludes the first part of our paper. To summarize: our main results
consist in the justification of the relative entropy eq.(\ref{S[p|P]}) as
the uniquely natural functional to select the best approximations and the
derivation of a quantitative measure of the degree to which the various
trials are preferred, eq.(\ref{p(theta)}). The final result for the best
approximation is eq.(\ref{pbar}).

Next we illustrate how this abstract formalism is used in a specific example.

\section{Simple fluids}

\label{sec3} Here we collect some necessary background material on simple
fluids.

We consider a simple fluid composed of $N$ single atom molecules described
by the Hamiltonian 
\begin{equation}
H(q_{N})=\sum\limits_{i=1}^{N}\,\frac{p_{i}^{2}}{2m}+U\quad \text{with}\quad
U=\sum\limits_{i>j}^{N}u(r_{ij})\,,  \label{Actual-H}
\end{equation}%
where $q_{N}=\{p_{i},r_{i};\;i=1,...,N\}$ and the many-body interactions are
approximated by a pair interaction, $u(r_{ij})$ where $r_{ij}=\left\vert
r_{i}-r_{j}\right\vert $. The probability that the positions and momenta of
the molecules lie within the phase space volume $dq_{N}$ is given by
canonical distribution, and 
\begin{equation}
P(q_{N})\,dq_{N}=\frac{1}{Z}e^{-\beta H(q_{N})\,}\,dq_{N}~,
\label{exact dist}
\end{equation}%
where 
\begin{equation}
dq_{N}=\frac{1}{N!\,h^{3N}}\prod\limits_{i=1}^{N}d^{3}p_{i}d^{3}r_{i}
\end{equation}%
and 
\begin{equation}
Z=\int dq_{N}\text{ }e^{-\beta H(q_{N})}.
\end{equation}%
For fluids dominated by pair interactions most thermodynamic quantities of
interest can be written in terms of the one- and two-particle density
distributions 
\begin{equation}
n(r)=\langle \hat{n}(r)\rangle \quad \text{and}\quad n^{\left( 2\right)
}(r_{1},r_{2})=\langle \hat{n}^{\left( 2\right) }(r_{1},r_{2})\rangle
\end{equation}%
where 
\begin{equation}
\hat{n}(r)=\sum\limits_{i}\,\delta (r-r_{i})
\end{equation}%
and 
\begin{equation}
\hat{n}^{\left( 2\right) }(r_{1},r_{2})=\sum\limits_{i,j(i\neq j)}\,\delta
(r_{1}-r_{i})\,\delta (r_{2}-r_{j})\;.
\end{equation}%
It is convenient to introduce the two-particle correlation function 
\begin{equation}
g(r_{1},r_{2})=\frac{n^{\left( 2\right) }(r_{1},r_{2})}{n(r_{1})n(r_{2})}~,
\end{equation}%
which measures the extent to which the structure of liquids deviates from
complete randomness. If the fluid is homogeneous and isotropic $n(r)=\rho
=N/V$ and $g(r_{1},r_{2})=g(|r_{1}-r_{2}|)=g(r)$ where $\rho $ is the bulk
density and $g(r)$ is the radial distribution function (RDF). Then, the
pressure is given by 
\begin{equation}
\frac{PV}{Nk_{B}T}=1-\frac{\beta \rho }{6}\int d^{3}r\,r\frac{du\left(
r\right) }{dr}g\left( r\right) \,,  \label{eq of state}
\end{equation}%
where $\beta \overset{\text{def}}{=}1/k_{B}T$. \cite{BarkerHenderson76}-\cite%
{Kalikmanov02}

The difficulty, of course, is that it is very difficult to calculate $g(r)$
from the ``exact'' distribution Eq.(\ref{exact dist}) and we need to find an
approximation that is calculable and still includes the two features of the
interaction potential $u$ that are relevant for explaining most fluid
properties: the strong repulsion at short distances and the weak attraction
at long distances.

To account for the short-distance repulsion we consider a family of trials
composed by distributions that describe a gas of hard spheres of diameter $%
r_{d}$. For each $r_{d}$ the Hamiltonian is 
\begin{equation}
H_{hs}(q_{N}\left\vert r_{d}\right. )=\sum\limits_{i=1}^{N}\,\frac{p_{i}^{2}%
}{2m}+U_{hs}  \label{Hhs}
\end{equation}%
with 
\begin{equation}
U_{hs}=\sum\limits_{i>j}^{N}u_{hs}(r_{ij}|r_{d})~,
\end{equation}%
where 
\begin{equation}
u_{hs}(r\left\vert r_{d}\right. )=\left\{ 
\begin{array}{ccc}
0 & \text{for} & r\geq r_{d} \\ 
\infty & \text{for} & r<r_{d}%
\end{array}%
\right.
\end{equation}%
and the corresponding probability distribution is 
\begin{equation}
P_{hs}(q_{N}\left\vert r_{d}\right. )=\frac{1}{Z_{hs}}e^{-\beta
H_{hs}(q_{N}\left\vert r_{d}\right. )}\,.  \label{Phs}
\end{equation}%
The partition function and the free energy $F_{hs}(T,V,N\left\vert
r_{d}\right) $ are 
\begin{equation}
Z_{hs}=\int dq_{N}\,\text{\ }e^{-\beta H_{hs}(q_{N}\left\vert r_{d}\right.
)}\,\overset{\text{def}}{=}e^{-\beta F_{hs}(T,V,N\left\vert r_{d}\right.
)}\,.
\end{equation}%
Two objections that can be raised to the choosing $P_{hs}(q_{N}|r_{d})$ as
trials are, first, that they do not take the long-range interactions into
account; and second, that the actual short range potential is not that of
hard spheres. These are points to which we will return later. A third
objection, and this is considerably more serious, is that the exact
hard-sphere RDF is not known. However, it can be calculated within the
approximation of Percus and Yevick (PY) for which there exists an exact
analytical solution (\cite{PercusYevick58}, \cite{Wertheim63}) which is
reasonably simple and in good agreement with numerical simulations over an
extended range of temperatures and densities, except perhaps at high
densities. There are several successful proposals \cite{Bravo91} to improve
upon the PY RDF but they also represent an additional level of complication.
We feel that the simpler PY RDF is sufficiently accurate for our current
objective -- to illustrate the application and study the broad features of
the ME approach. We will therefore assume that for our purposes the $P_{hs}$
are sufficiently tractable distributions.

The PY RDF can be written in terms of the Laplace transform of $%
rg_{hs}(r\left\vert r_{d}\right. )$ \cite{Wertheim63}, 
\begin{eqnarray}
G(s) &=&\int\limits_{0}^{\infty }dx~xg_{hs}(xr_{d}|r_{d})e^{-sx}  \notag \\
&=&\frac{sL(s)}{12\eta \left[ L(s)+M(s)e^{s}\right] },  \label{G(s)}
\end{eqnarray}%
where $x$ is a dimensionless variable $x=r/r_{d}$, 
\begin{equation}
L(s)=12\eta \left[ \left( 1+\frac{1}{2}\eta \right) s+\left( 1+2\eta \right) %
\right] ,
\end{equation}%
\begin{eqnarray}
M(s) &=&\left( 1-\eta \right) ^{2}s^{3}+6\eta \left( 1-\eta \right) s^{2} 
\notag \\
&+&18\eta ^{2}s-12\eta \left( 1+2\eta \right) ,
\end{eqnarray}%
and $\eta $ is the packing fraction, 
\begin{equation}
\eta \overset{\text{def}}{=}\frac{1}{6}\pi \rho r_{d}^{3}\quad \text{with}%
\quad \rho =\frac{N}{V}~.  \label{eta}
\end{equation}%
The RDF $g_{hs}(r\left\vert r_{d}\right. )$ is obtained from the inverse
transform using residues \cite{Throop65}.

The equation of state can then be computed in two alternative ways, either
from the ``pressure'' equation or from the ``compressibility'' equation but,
since the result above for $g_{hs}(r\left\vert r_{d}\right. )$ is not exact,
the two results do not agree. It has been found that better agreement with
simulations and with virial coefficients is obtained taking an average of
the two results with weights 1/3 and 2/3 respectively. The result is the
Carnahan-Starling equation of state, \cite{BarkerHenderson76}-\cite%
{Kalikmanov02} 
\begin{equation}
\left( \frac{PV}{Nk_{B}T}\right) _{hs}=\frac{1+\eta +\eta ^{2}-\eta ^{3}}{%
\left( 1-\eta \right) ^{3}}.  \label{HS-P}
\end{equation}%
The free energy, derived by integrating the equation of state, is 
\begin{equation}
F_{hs}(T,V,N\left\vert r_{d}\right. )=Nk_{B}T\left[ -1+\ln \rho \Lambda ^{3}+%
\frac{4\eta -3\eta ^{2}}{\left( 1-\eta \right) ^{2}}\right] ,\text{ }
\label{HS-Free energy}
\end{equation}%
where $\Lambda =(2\pi \hbar ^{2}/mk_{B}T)^{1/2}$, and the entropy is 
\begin{equation}
S_{hs}=-\left( \frac{\partial F_{hs}}{\partial T}\right) _{N,V}=\frac{F_{hs}%
}{T}+\frac{3}{2}Nk_{B}.  \label{HS-S}
\end{equation}%
It must be remembered that these expressions are not exact. They are
reasonable approximations for all densities up to almost crystalline
densities (about $\eta \approx 0.5$). However, they fail to predict the
face-centered-cubic phase when $\eta $ is in the range from $0.5$ up the
close-packing value of $0.74$.

\section{The optimal hard-sphere diameter}

\label{sec4} As discussed in section ~\ref{sec2}, the trial $%
P_{hs}(q_{N}|r_{d})$ that is ``closest'' to the ``exact'' $P(q_{N})$ is
found by maximizing the relative entropy, 
\begin{equation}
S\left[ P_{hs}|P\right] =-\int dq_{N}\,P_{hs}(q_{N}|r_{d})\log \frac{%
P_{hs}(q_{N}|r_{d})}{P(q_{N})}\text{ }\leq 0\text{.}
\end{equation}%
Substituting Eqs.(\ref{exact dist}) and (\ref{Phs}) we obtain 
\begin{equation}
S\left[ P_{hs}|P\right] =\beta \left[ F-F_{hs}-\langle U-U_{hs}\rangle _{hs}%
\right] \leq 0\,,  \label{S1}
\end{equation}%
where $\langle \cdots \rangle _{hs}$ is computed over the hard-sphere
distribution $P_{hs}(q_{N}|r_{d})$. Eq.(\ref{S1}) can be rewritten as 
\begin{equation}
F\leq F_{U}\overset{\text{def}}{=}F_{hs}+\langle U-U_{hs}\rangle _{hs}\text{
,}
\end{equation}%
which shows that maximizing $S\left[ P_{hs}|P\right] $ is equivalent to
minimizing $F_{U}$ over all diameters $r_{d}$. Thus, the variational
approximation to the free energy is 
\begin{equation}
F\left( T,V,N\right) \approx F_{U}(T,V,N\left\vert r_{m}\right. )\overset{%
\text{def}}{=}\min_{r_{d}}~F_{U}(T,V,N\left\vert r_{d}\right. )\,,
\label{mini-Fu}
\end{equation}%
where $r_{m}$ is the optimal diameter.

To calculate $F_{U}$ use 
\begin{eqnarray}
\langle U-U_{hs}\rangle _{hs} &=&\frac{1}{2}\int d^{3}rd^{3}r^{\prime }\text{
}n_{hs}^{\left( 2\right) }(r,r^{\prime })  \notag \\
&&\left[ u(r-r^{\prime })-u_{hs}(r-r^{\prime }|r_{d})\right] ~,
\end{eqnarray}%
where $n_{hs}^{\left( 2\right) }(r,r^{\prime })=\langle \hat{n}^{\left(
2\right) }(r,r^{\prime })\rangle _{hs}$. But $u_{hs}(r-r^{\prime }|r_{d})=0$
for $\left\vert r-r^{\prime }\right\vert \geq r_{d}$ while $n_{hs}^{\left(
2\right) }(r,r^{\prime })=0$ for $\left\vert r-r^{\prime }\right\vert \leq
r_{d}$, therefore 
\begin{equation}
F_{U}=F_{hs}+\langle U\rangle _{hs}  \label{F_U}
\end{equation}%
with 
\begin{equation}
\langle U\rangle _{hs}=\frac{1}{2}N\rho \int d^{3}r\,u(r)g_{hs}(r\left\vert
r_{d}\right. ),
\end{equation}%
where we have assumed that the fluid is isotropic and homogeneous, $%
n_{hs}^{\left( 2\right) }(r,r^{\prime })=n_{hs}^{\left( 2\right)
}(\left\vert r-r^{\prime }\right\vert )$, and introduced the hard-sphere
radial distribution function 
\begin{equation}
g_{hs}(r\left\vert r_{d}\right. )\overset{\text{def}}{=}\frac{n_{hs}^{\left(
2\right) }(r)}{\rho ^{2}}.
\end{equation}%
Notice that the approximation does not consist of merely replacing the exact
free energy $F$ by a hard-sphere free energy $F_{hs}$ which does not include
the effects of long range attraction; $F$ is approximated by $F_{U}(r_{m})$
which includes attraction effects through the $\langle U\rangle _{hs}$ term
in eq.(\ref{F_U}). This addresses the first of the two objections mentioned
earlier: the real fluid with interactions given by $u$ is not being replaced
by a hard-sphere fluid with interactions given by $u_{hs}$; it is just the
probability distribution that is being replaced in this way. The internal
energy is approximated by $\langle H\rangle _{hs}=\frac{3}{2}Nk_{B}T+\langle
U\rangle _{hs}$ and not by $\langle H_{hs}\rangle _{hs}=\frac{3}{2}Nk_{B}T$.

To calculate $\langle U\rangle _{hs}$ it is convenient to write it in terms
of $V(s)$, the inverse Laplace transform of $ru(r)$, 
\begin{equation}
xu(xr_{d})=\int\limits_{0}^{\infty }ds~V(s)e^{-sx}.
\end{equation}%
For example, for a Lennard-Jones potential, 
\begin{equation}
u(r)=4\varepsilon \left[ \left( \frac{\sigma }{r}\right) ^{12}-\left( \frac{%
\sigma }{r}\right) ^{6}\right] ,  \label{LJ Pot}
\end{equation}%
we have 
\begin{equation}
V(s)=4\varepsilon \left[ \left( \frac{\sigma }{r_{d}}\right) ^{12}\frac{%
s^{10}}{10!}-\left( \frac{\sigma }{r_{d}}\right) ^{6}\frac{s^{4}}{4!}\right]
.
\end{equation}%
Then, using eq.(\ref{F_U}) and eq.(\ref{G(s)}) gives 
\begin{equation}
\langle U\rangle _{hs}=12N\eta \int\limits_{0}^{\infty }ds\text{~}V(s)G(s).
\label{F_U-LT}
\end{equation}%
Finally, it remains to minimize $F_{U}$ in eq.(\ref{F_U}) to determine the
optimal diameter $r_{m}$. This is done numerically; an explicit example for
Argon is calculated in section ~\ref{sec5}.\ 

As mentioned earlier the problem with the approach outlined above is that it
fails to take the softness of the repulsive core into account. This flaw is
manifested in a less satisfactory prediction of thermodynamic variables at
high temperatures, and also, as will be shown by the numerical calculations
in section ~\ref{sec6}, in a poor prediction of the radial distribution
function.

\section{Beyond the Variational Principle}

\label{sec5} The ME method as pursued in the last section has led us to
determine an optimal value of the hard-sphere diameter. Next we consider
whether the correct choice should have been some other value $%
r_{d}=r_{m}+\delta r$ rather than the optimal $r_{d}=r_{m}$. As discussed in
section ~\ref{sec2-3} this is a question about the probability of $r_{d}$, $%
P_{d}(r_{d})$. Thus, we are uncertain not just about $q_{N}$ but also about $%
r_{d}$ and what we actually seek is the joint probability of $q_{N}$ and $%
r_{d}$, $P_{J}(q_{N},r_{d})$. Once this joint distribution is obtained our
best assessment of the distribution of $q_{N}$ is given by the marginal over 
$r_{d}$, 
\begin{eqnarray}
\bar{P}_{hs}(q_{N}) &\overset{\text{def}}{=}&\int dr_{d}\text{ }%
P_{J}(q_{N},r_{d})  \notag \\
&=&\int dr_{d}\text{~}P_{d}(r_{d})P_{hs}(q_{N}|r_{d}).\text{ }
\label{Pbar(q)}
\end{eqnarray}%
By recognizing that diameters other than $r_{m}$ are not ruled out and that
a more honest representation is an average over all hard-sphere diameters we
are effectively replacing the hard spheres by a soft-core potential.

However, we should emphasize that the distribution over the hard-sphere
diameters $P_{d}(r_{d})$ is not being introduced in an ad hoc way in order
to account for the softness of the LJ potential. The introduction of $%
p(\theta )$ in general (section ~\ref{sec2-3}), and of $P_{d}(r_{d})$ in
particular, are mandated by the ME method. The distribution $p(\theta )$
would also arise if one were trying to find an ME approximation to other
problems which do not involve hard spheres at all.

The distribution of diameters is given by eq.(\ref{p(theta)}) 
\begin{equation}
P_{d}(r_{d})dr_{d}=\frac{e^{S\left[ P_{hs}|P\right] }}{\zeta }\gamma
^{1/2}\left( r_{d}\right) dr_{d}=\frac{e^{-\beta F_{U}}}{\zeta _{U}}\gamma
^{1/2}\left( r_{d}\right) dr_{d},  \label{Pd(rd)}
\end{equation}%
where $S\left[ P_{hs}|P\right] =\beta \left( F-F_{U}\right) ,$ the partition
functions $\zeta $ and $\zeta _{U}$ are given by 
\begin{equation}
\zeta =e^{\beta F}\zeta _{U}\quad \text{with}\quad \zeta _{U}=\int dr_{d}%
\text{ }\gamma ^{1/2}\left( r_{d}\right) e^{-\beta F_{U}},
\end{equation}%
and the natural distance $d\ell ^{2}=\gamma (r_{d})dr_{d}^{2}$ in the space
of $r_{d}$s is given by the Fisher-Rao metric, 
\begin{equation}
\gamma (r_{d})=\int dq_{N}\,P_{hs}(q_{N}\left\vert r_{d}\right. )\left( 
\frac{\partial \log P_{hs}(q_{N}\left\vert r_{d}\right. )}{\partial r_{d}}%
\right) ^{2}.  \label{FR-metric}
\end{equation}%
%
%
The remaining problem in the above equations is the calculation of the
Fisher-Rao measure $\gamma ^{1/2}$ and this is conveniently done by
considering the entropy of $P_{hs}(q_{N}\left\vert r_{d}^{\prime }\right. )$
relative to $P_{hs}(q_{N}\left\vert r_{d}\right. )$. A straightforward
differentiation shows that 
\begin{equation}
-\left. \frac{\partial ^{2}S\left[ P_{hs}(\cdot \left\vert r_{d}^{\prime
}\right. )\left\vert P_{hs}(\cdot \left\vert r_{d}\right. )\right. \right] }{%
\partial r_{d}^{\prime 2}}\right\vert _{r_{d}^{\prime }=r_{d}}=\gamma
(r_{d}).  \label{det g}
\end{equation}%
Substituting the distributions $P_{hs}(q_{N}\left\vert r_{d}^{\prime
}\right. )$ and $P_{hs}(q_{N}\left\vert r_{d}\right. )$ gives 
\begin{equation}
S\left[ P_{hs}(.|r_{d}^{\prime })|P_{hs}(.\left\vert r_{d}\right. )\right]
=\beta \left[ \left. F_{hs}\right\vert _{r_{d}^{\prime }}^{r_{d}}-\langle
\left. U_{hs}\right\vert _{r_{d}^{\prime }}^{r_{d}}\rangle _{r_{d}^{\prime }}%
\right] ,
\end{equation}%
where $\langle \cdots \rangle _{r_{d}^{\prime }}$ is the average over $%
P_{hs}(q_{N}|r_{d}^{\prime })$. As we argued above eq.(\ref{F_U}) the
expectation of the potential energy $\langle U_{hs}\left( r_{d}^{\prime
}\right) \rangle _{r_{d}^{\prime }}$ vanishes because the product $%
u(r\left\vert r_{d}^{\prime }\right. )g_{hs}(r\left\vert r_{d}^{\prime
}\right. )$ vanishes for both $r<r_{d}^{\prime }$ and $r>r_{d}^{\prime }$.
Similarly, $\langle U_{hs}\left( r_{d}\right) \rangle _{r_{d}^{\prime }}=0$
when $r_{d}^{\prime }>r_{d}$. However, when $r_{d}^{\prime }<r_{d}$ the
expectation $\langle U_{hs}\left( r_{d}\right) \rangle _{r_{d}^{\prime }}$
diverges, $S$ is not defined and eq.(\ref{det g}) is not applicable. We can
argue our way out of this quandary by pointing out that the divergence is a
consequence of the unphysical idealization involved in taking a hard-sphere
potential seriously. For more realistic continuous potentials the distance
between $r_{d}^{\prime }=r_{d}+dr_{d}$ and $r_{d}$ is the same as the
distance between $r_{d}^{\prime }=r_{d}-dr_{d}$ and $r_{d}$. We can then
always choose $r_{d}^{\prime }\geq r_{d}$ and define $\gamma (r_{d})$ in eq.(%
\ref{det g}) as the limit $r_{d}^{\prime }=r_{d}+0^{+}$. Then, using eq.(\ref%
{HS-Free energy}), we have 
\begin{eqnarray}
\gamma (r_{d}) &=&\beta \left. \frac{\partial ^{2}F_{hs}\left( r_{d}^{\prime
}\right) }{\partial r_{d}^{\prime 2}}\right\vert _{r_{d}^{\prime
}=r_{d}+0^{+}}\text{ }  \notag \\
&=&N\pi \rho r_{d}\frac{4+9\eta -4\eta ^{2}}{\left( 1-\eta \right) ^{4}}~.
\label{det g-exact}
\end{eqnarray}

To summarize, the distribution of diameters $P_{d}(r_{d})$ is given by eq.(%
\ref{Pd(rd)}) with $F_{U}$ given by Eqs.(\ref{F_U}, \ref{HS-Free energy}, %
\ref{F_U-LT}) and $\gamma $ given by (\ref{det g-exact}). Our best
approximation to the ``exact'' $P(q_{N})$ is the $\bar{P}_{hs}(q_{N})$ given
in eq.(\ref{Pbar(q)}). The corresponding best approximation to the radial
distribution function is

\begin{equation}
\bar{g}_{hs}(r)=\int dr_{d}\text{~}P_{d}(r_{d})g_{hs}(r\left\vert
r_{d}\right. )\text{ . }  \label{ghs bar}
\end{equation}

Since $\bar{g}_{hs}(r)$ takes into account soft-core effects while $%
g_{hs}(r\left\vert r_{m}\right. )$ does not, we expect that it will lead to
improved estimates for all other thermodynamic quantities. However, there is
a problem. Since the free energy $F_{U}$ is an extensive quantity, $%
F_{U}\propto N$, for large $N$ the distribution $P_{d}(r_{d})\sim \exp
-\beta F_{U}$ is very sharply peaked at the optimal diameter $r_{m}$. This
result must be interpreted with care: when choosing a \emph{single} optimal
diameter for a macroscopic fluid sample we find that ME confers overwhelming
probability to the optimal value. This is not surprising. The same thing
happens when we calculate the global temperature or density of a macroscopic
sample: standard textbook results predict that fluctuations about the
expected value are utterly negligible. And yet fluctuations can be
important. For example, for small fluid samples, or when we consider the
local behavior of the fluid, fluctuations are not merely observable but can
be large. Local fluctuations can be appreciable while global fluctuations
remain negligible. The question then, is whether these local fluctuations
are relevant to the particular quantities we want to calculate. We argue
that they are.

The radial distribution function $g(r)$ is the crucial quantity from which
all other thermodynamic variables are computed. But from its very definition
as the probability that given an atom at a certain place another atom will
be found at a distance $r$, it is clear that $g(r)$ refers to purely local
behavior and should be affected by local fluctuations. To the extent that
the optimal $r_{m}$ depends on temperature and density we expect that local
temperature and/or density fluctuations would induce local diameter
fluctuations as well.

The extended analysis in this section does not yet allow us to pursue the
question of local diameter fluctuations in a satisfactory manner. As we
mentioned earlier the ME method is quite rigid in that the only freedom it
allows is the choice of trial distributions. A proper analysis of diameter
fluctuations would require enlarging the family of trial distributions to
allow for spatial inhomogeneities in the diameters of the hard spheres. It
may be worthwhile spelling out one possible such enlargement. We could
imagine a trial model where the molecules are hard spheres with a diameter
that depends on their location $r_{d}(\vec{r})$. As a molecule moves around
its diameter shrinks and expands according to a prescribed function $r_{d}(%
\vec{r})$ which thus acquires the character of an external ``field''. To
each possible choice of the diameter field $r_{d}(\vec{r})$ there
corresponds one trial distribution. This means that instead of a
one-dimensional family of trials labelled by the single parameter $r_{d}$ we
would have to deal with an infinite-dimensional family labeled by the fields 
$r_{d}(\vec{r})$. One should remark that these trial distributions do not
describe any ``physical system but this in itself is not a problem. The ME
method does not attempt to approximate one physical system by another
physical, albeit idealized system; it just attempts to approximate one
mathematical distribution by another; there is no requirement that the
latter be interpretable in terms of physically realizable Hamiltonians. The
real problem, of course, is that these inhomogeneous trial models are not
(at present) sufficiently tractable. However, we could divide the fluid into
mesoscopic cells and consider trial models where the diameters $r_{d}(i)$
are uniform within each cell $i$, which allows a local Percus-Yevick
approximation. The ME method would then be applied to determine not only the
distribution of diameters within each cell but also the optimal size of the
cells. This is a development we plan to pursue in future work.

For the purpose of this paper, however, we can quickly estimate the effects
of local fluctuations by pointing out that the size of the region (i.e., the
size of the cell) that is locally relevant to the calculation of $g(r)$
contains an effective number of atoms $N_{\text{eff}}$ that can be estimated
directly from a feature of the exact RDF that turns out to be known. (What
we are doing is making the best use of information that happens to be
available, which is quite in the spirit of our information theory-maximum
entropy approach.)

The basic idea is intuitively simple: at very short distances the form of
the true, exact RDF\ $g\left( r\right) $ is dominated by the repulsive part
of the potential. If the size of the molecule is given by the Lennard-Jones
parameter $\sigma $, eq.(\ref{LJ Pot}), the asymptotic form of $g(r)$ is
given by 
\begin{equation}
g\left( r\right) \rightarrow e^{-\beta u\left( r\right) }\quad \text{%
for\quad }r\ll \sigma ~.\text{ }  \label{estimaterule}
\end{equation}
For a sufficiently dilute gas $g\left( r\right) \approx e^{-\beta u\left(
r\right) }$ holds for all distances $r$; for dense fluids eq.(\ref%
{estimaterule}) is valid only for $r\rightarrow 0$. (This follows from a
clever trick due to Percus which allows one to write an exact expression for
the two-particle distribution function for a fluid in terms of the single
particle density of the same fluid placed in a suitable external potential 
\cite{Percus62}.) A fluid of hard spheres gives $g(r|r_{d})=0$ for $r<r_{d}$
and cannot reproduce the behavior (\ref{estimaterule}). However, once we
recognize that we can use a statistical mixture, eq.(\ref{Pbar(q)}), we can
tune the size $N_{\text{eff}}$ of the cell and thereby change the width of $%
P_{d}\left( r_{d}\right) $ so that the radial distribution function $\bar{g}%
_{hs}\left( r\right) $ of (\ref{ghs bar}) reproduces the known
short-distance behavior. This we proceed to do in the next section.

\section{An example: Lennard-Jones ``Argon''}

\label{sec6} One of the difficulties in testing theories about fluids
against experimental data is that it is not easy to see whether
discrepancies are to be blamed to a faulty approximation or to a wrong
intermolecular potential. This is why theories are normally tested against
molecular dynamics numerical simulations where there is control over the
intermolecular potential. In this section we compare ME results against
simulation results \cite{Verlet68} for a fluid of monatomic molecules
interacting through a Lennard-Jones potential, eq.(\ref{LJ Pot}). The
parameters $\varepsilon $ and $\sigma $ (the depth of the well, $u|_{\min
}=-\varepsilon $, and the radius of the repulsive core, $u(\sigma )=0$,
respectively) are chosen to model Argon: $\varepsilon =1.03\times 10^{-2}$ $%
\text{eV}$ and $\sigma =3.405 $. 
{\AA}%
%
%
%

\subsection{The free energy $F_{U}$}

\label{sec6-1} 
\begin{figure}[ht]
\centering
\begin{picture}(0,150)(0,0)
\put(0,150){(A)}
\end{picture}
\includegraphics[width=2.0in,height=2.0in]{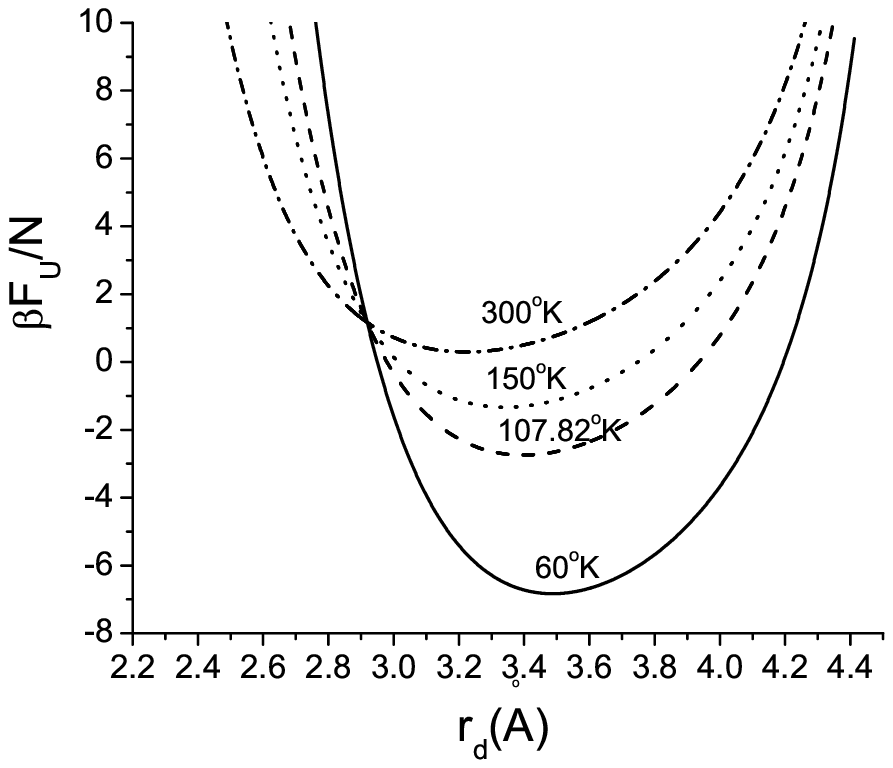} 
\begin{picture}(5,150)(0,0)
\put(5,150){(B)}
\end{picture}
\includegraphics[width=2.0in,height=2.0in]{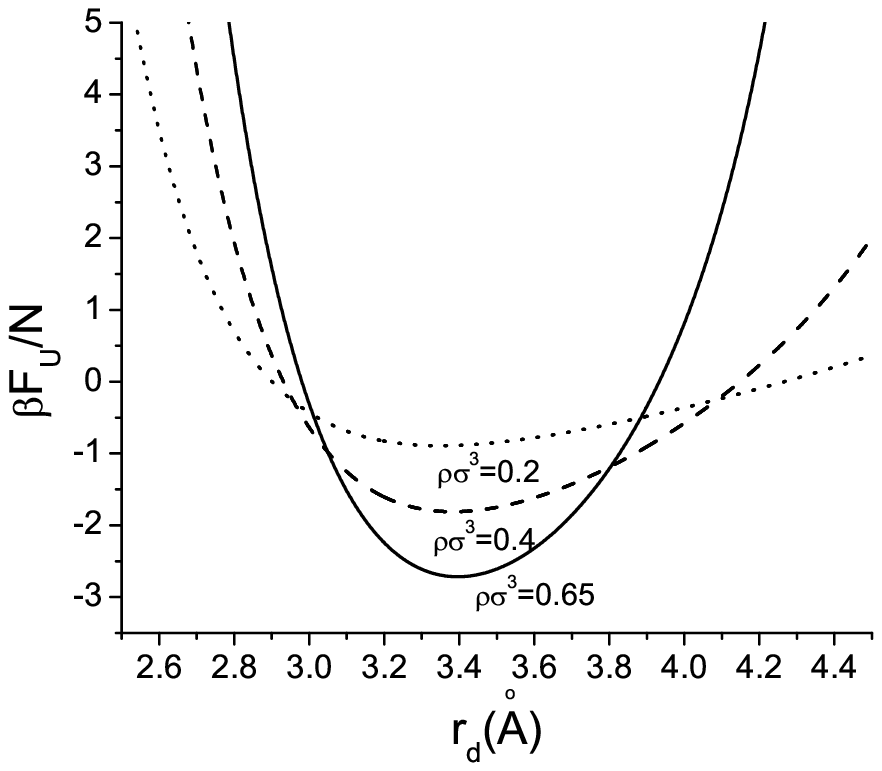}
\caption{(A): The free energy $F_{U}$ as a function of hard-sphere diameter $%
r_{d}$ for Argon at a density of $\protect\rho \protect\sigma ^{3}=0.65$ for
different temperatures. The best $r_{d}$ is that which minimizes $F_{U}$.
(B): $F_{U}$ as a function of $r_{d}$ for Argon at $T=107.82$ $\text{K}$ for
different densities.}
\label{fig1}
\end{figure}

Figure \ref{fig1}.(A) shows the free energy $F_{U}/Nk_{B}T$
as a function of hard-sphere diameter $r_{d}$ for Argon at a fixed density
of $\rho \sigma ^{3}=0.65$ for different temperatures. Figure \ref{fig1}.(B)
shows $F_{U}/Nk_{B}T $ as a function of $r_{d}$ for several densities at
fixed $T=107.82$ $\text{K}$. Since the critical point for Argon is at $%
T_{c}=150.69$ $\text{K}$ and $\rho _{c}\sigma ^{3}=0.33$ all these curves,
except that at $300$ $\text{K}$, lie well within the liquid phase. The
increase of $F_{U}/Nk_{B}T$ for high values of $r_{d}$ is due to short range
repulsion between the hard spheres described by $F_{hs}/Nk_{B}T$. The
increase for low $r_{d}$ is due to the Lennard-Jones short-range repulsion
as described by $\langle U\rangle _{hs}/Nk_{B}T$.

The best $r_{d}$ is that which minimizes $F_{U}$ and depends both on
temperature and density. The best diameter decreases as the temperature
increases because atoms with higher energy can penetrate deeper into the
repulsive core. The dependence with density is less pronounced.

\subsection{The distribution of diameters $P_{d}(r_{d})$}

\label{sec6-2} 
\begin{figure}[ht]
\centering
\begin{picture}(0,150)(0,0)
\put(0,150){(A)}
\end{picture}
\includegraphics[width=2.0in,height=2.0in]{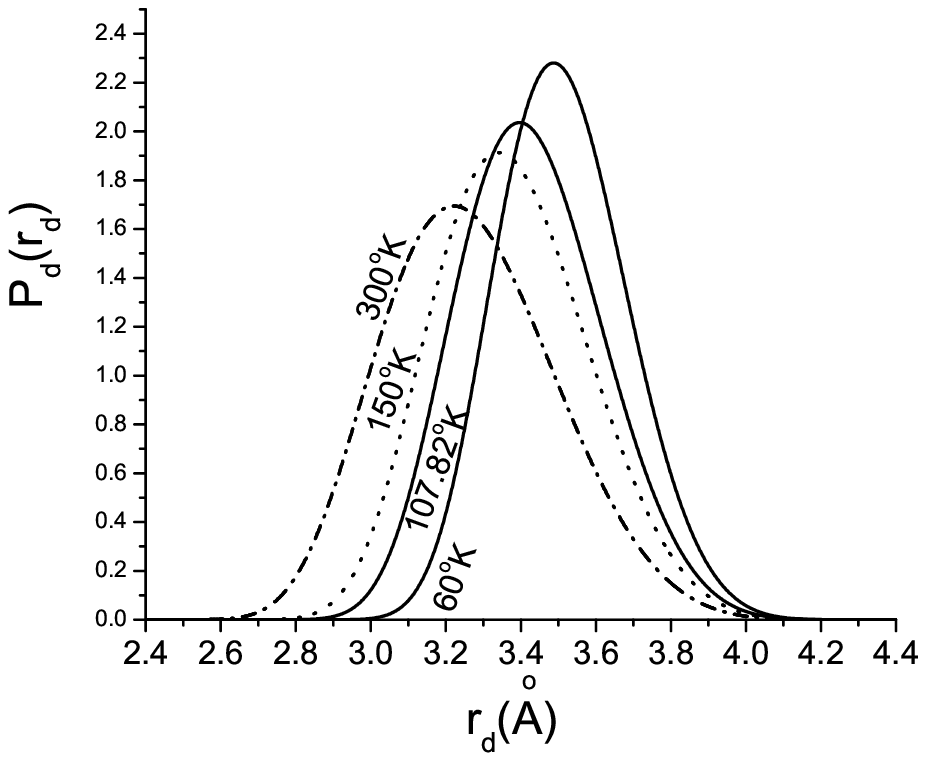} 
\begin{picture}(4,150)(0,0)
\put(4,150){(B)}
\end{picture}
\includegraphics[width=2.0in,height=2.0in]{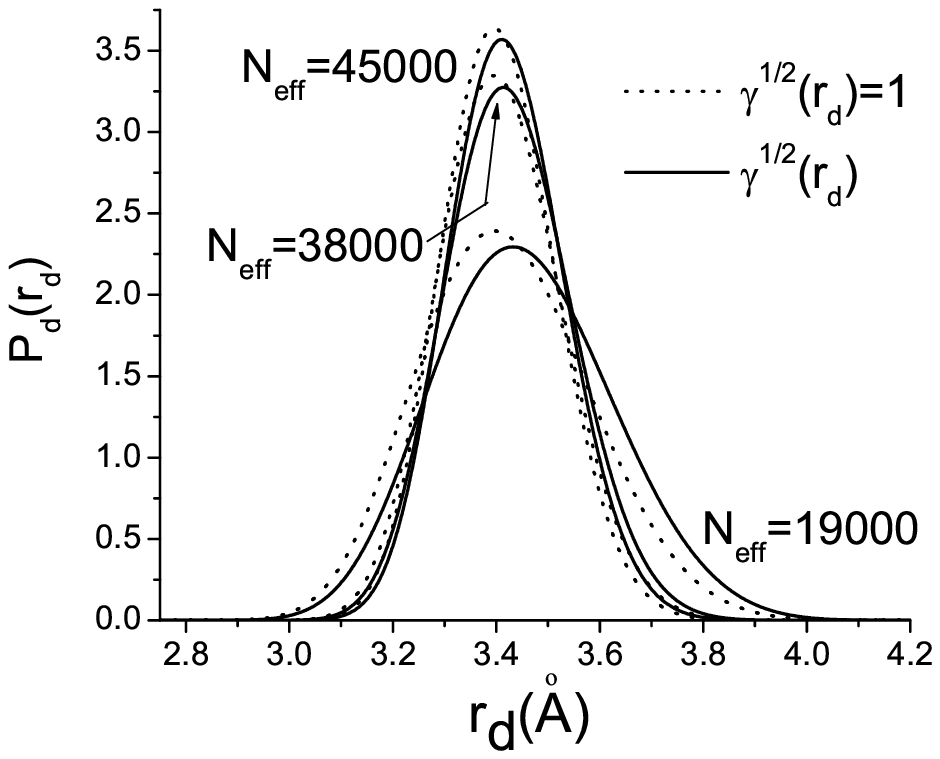}
\caption{(A): The distribution of hard-sphere diameters $r_{d}$ for Argon
for several temperatures at density $\protect\rho \protect\sigma ^{3}=0.65$
for $N_{\text{eff}}=13500$. (B): $P_{d}\left( r_{d}\right) $ for various $N_{%
\text{eff}}$ at $T=107.82$ $\text{K}$ and $\protect\rho \protect\sigma %
^{3}=0.65$. By setting $\protect\gamma ^{1/2}=1$ (dotted lines) we see that
the effect of the $\protect\gamma ^{1/2} $ factor is to cause a slight shift
of the distribution.}
\label{fig3}
\end{figure}

In section ~\ref{sec5} we argued that the effective number of
molecules that is relevant to the local structure of the fluid is not the
total number of molecules in the system $N$, but a smaller number, $N_{\text{%
eff}}$. In Fig. \ref{fig3}.(A) we plot the distribution of diameters $%
P_{d}(r_{d})$ for different temperatures, for a fixed fluid density of $\rho
\sigma ^{3}=0.65$, and for an arbitrarily chosen $N_{\text{eff}}=13500$. As
expected the distribution shifts to higher diameters as the temperature
decreases. Notice also that the distribution becomes narrower at lower
temperatures in agreement with the fact that a hard-sphere approximation is
better at low $T$ \cite{BarkerHenderson76}.

Figure \ref{fig3}.(B) shows that increasing $N_{\text{eff}}$ (with fixed
density $\rho $) decreases the width of $P_{d}(r_{d})$ (solid lines) and
induces a slight shift of the whole distribution. This is due to the
dependence $\sim (N_{\text{eff}}r_{d})^{1/2}$ of the Fisher-Rao measure $%
\gamma ^{1/2}\left( r_{d}\right) $ in Eq.(\ref{det g-exact}). Figure \ref%
{fig3}.(B) also explores the influence of $\gamma ^{1/2}\left( r_{d}\right) $
by comparing the actual distributions $P_{d}(r_{d})$ (solid lines) with the
distributions $e^{-\beta F_{U}\left( r_{d}\right) }$ (dotted lines) which
are obtained by setting $\gamma ^{1/2}=1$ in Eq.(\ref{Pd(rd)}). The effect
of $\gamma ^{1/2}$ is to shift the distribution slightly to higher $r_{d}$.


\subsection{The radial distribution function}

\label{sec6-3} 
\begin{figure}[ht]
\centering
\includegraphics[width=2.0in,height=2.0in]{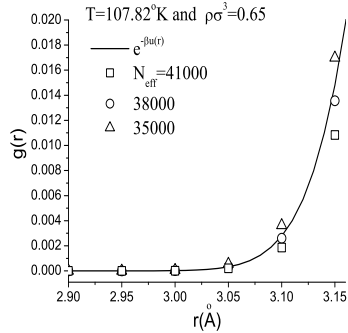}
\caption{ {Estimating $N_{\text{eff}}$ by requiring that $\bar{g}_{hs}(r)$
have the correct short-distace behavior $e^{-\protect\beta u(r)}$.}}
\label{fig5}
\end{figure}

We are finally ready to calculate the radial distribution $%
g(r)$ for Argon. We start by estimating the number of molecules $N_{\text{eff%
}}$ that are locally relevant; as explained earlier we choose $N_{\text{eff}%
} $ so that our best approximation $\bar{g}_{hs}\left( r\right) $, Eq. (\ref%
{ghs bar}), reproduces the known short-distance behavior $e^{-\beta u\left(
r\right) }$. We have found that the estimates for $N_{\text{eff}}$ need not
be very accurate but that they must be obtained for each value of the
temperature and density. In Fig. \ref{fig5} we show an example of the
short-distance behavior of $\bar{g}_{hs}$ for three values of $N_{\text{eff}%
} $ at $T=107.82$ $\text{K}$ and $\rho \sigma ^{3}=0.65$; using a chi-square
fit in the range from $r=2.9$ to $3.1$ $\text{\AA }$ the selected best value
of $N_{\text{eff}}$ is around 38000.

\begin{figure}[th]
\centering
\begin{picture}(0,150)(0,0)
\put(0,150){(A)}
\end{picture}
\includegraphics[width=2.0in,height=2.0in]{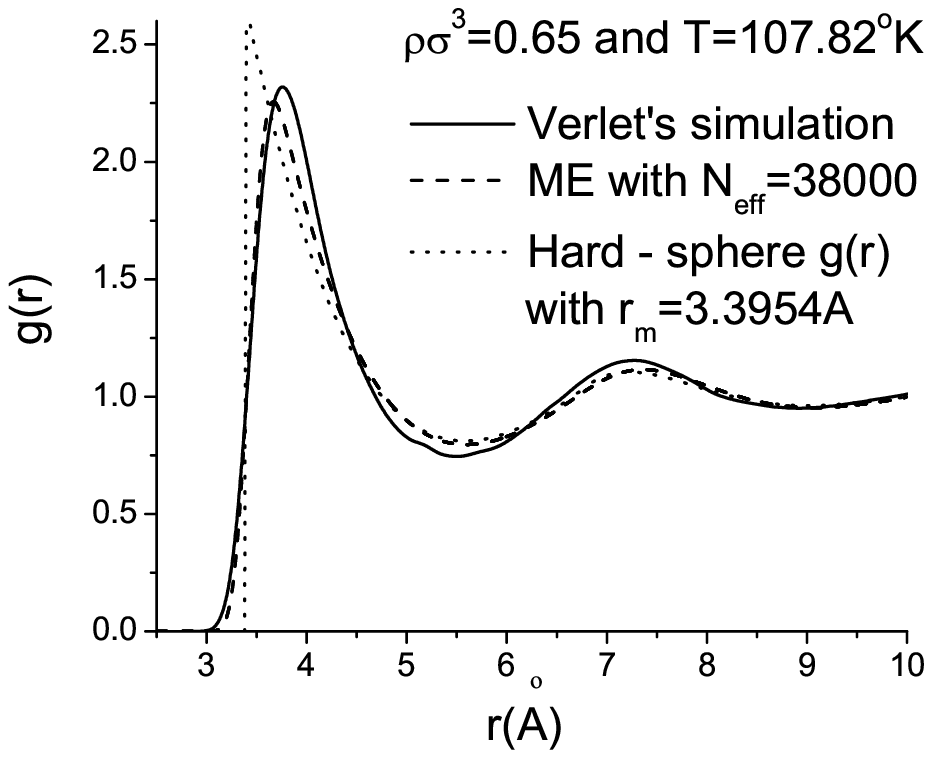} 
\begin{picture}(4,150)(0,0)
\put(0,150){(B)}
\end{picture}
\includegraphics[width=2.0in,height=2.0in]{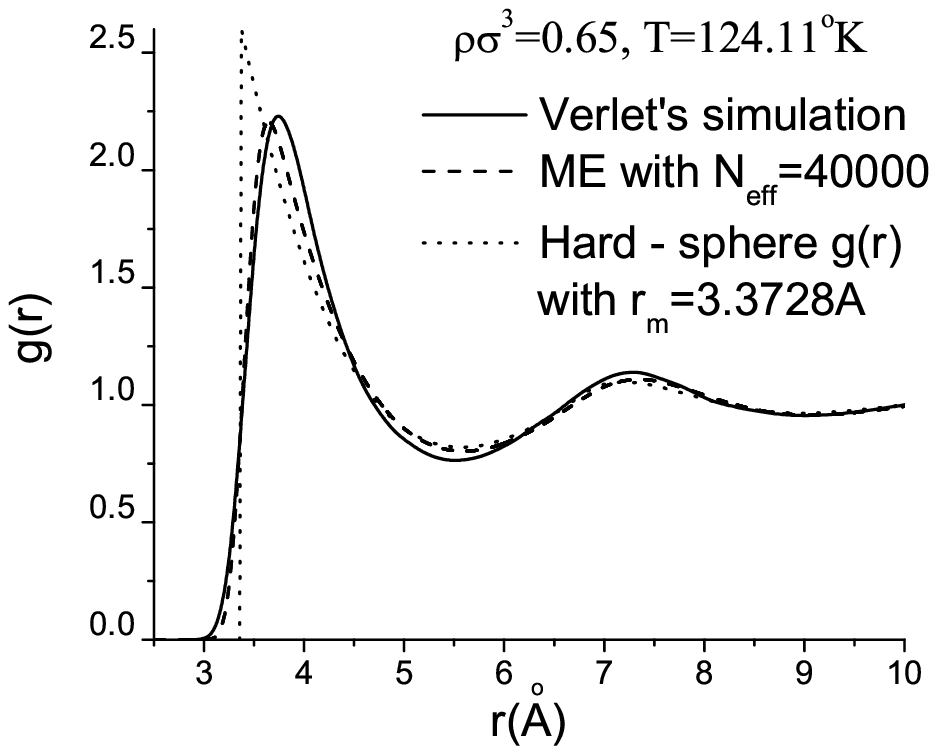}\newline
\begin{picture}(0,155)(0,0)
\put(0,155){(C)}
\end{picture}
\includegraphics[width=2.0in,height=2.0in]{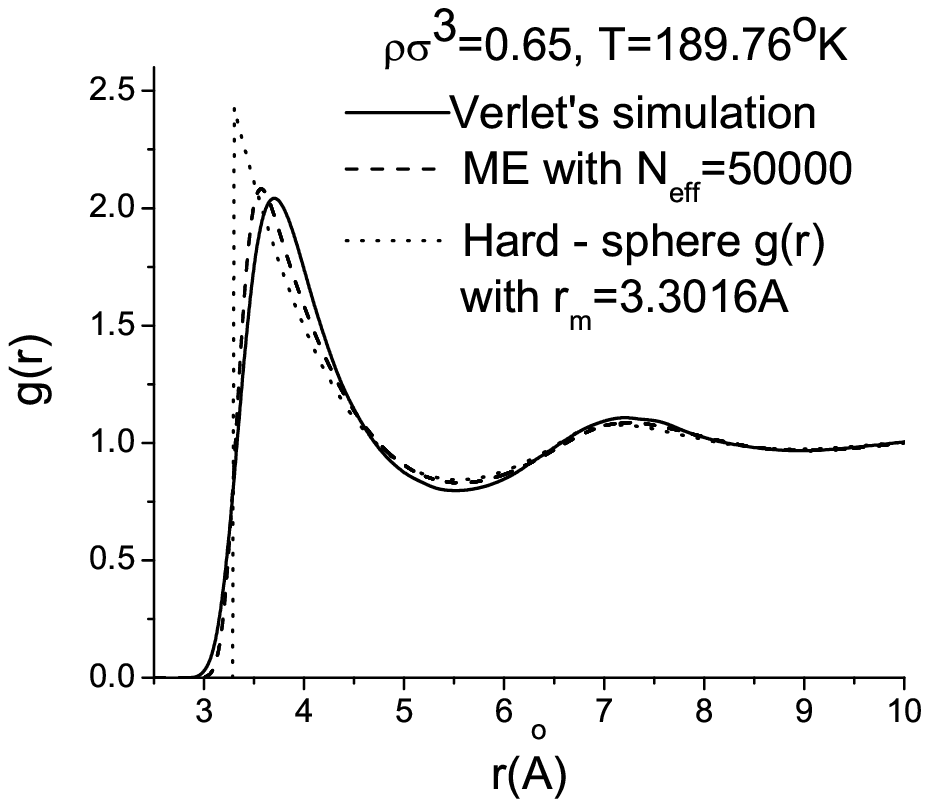} 
\begin{picture}(5,155)(0,0)
\put(0,155){(D)}
\end{picture}
\includegraphics[width=2.0in,height=2.0in]{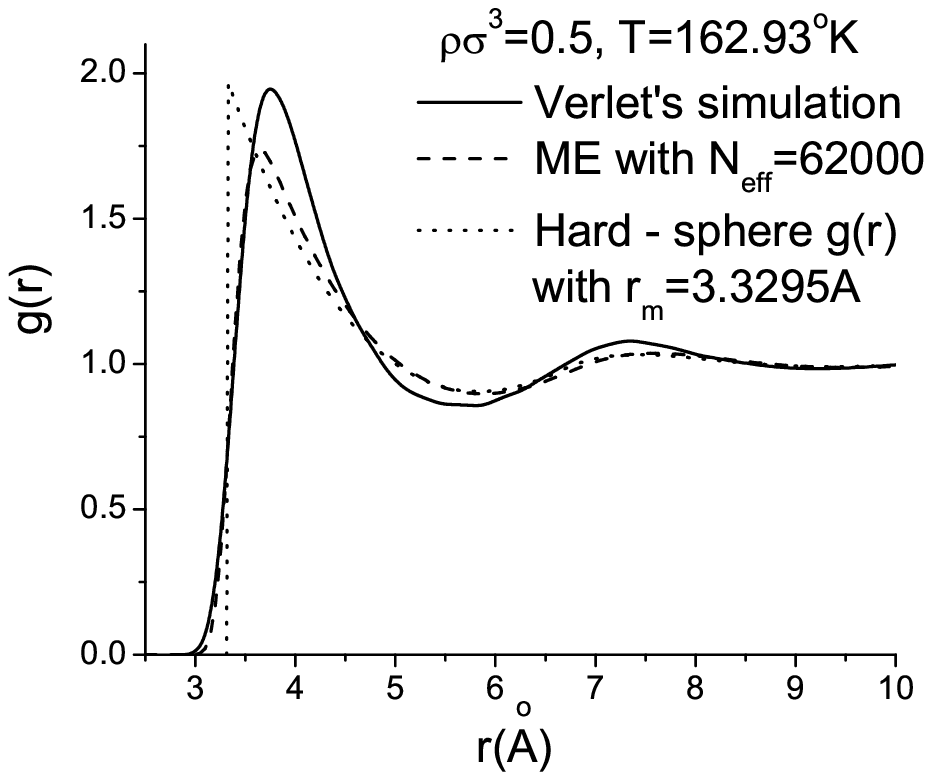}
\caption{The radial distribution function for (a) the hard-sphere fluid with
optimal diameter $r_{m}$; (b) Verlet's molecular dynamics simulation; and
(c) the improved ME analysis, for Argon at (A): density $\protect\rho 
\protect\sigma ^{3}=0.65$, temperature $T=107.82$ $\text{K}$, and effective
particle number $N_{\text{eff}}=38000$. (B): $\protect\rho \protect\sigma %
^{3}=0.65$, $T=124.11$ $\text{K}$, and $N_{\text{eff}}=40000$. (C): $\protect%
\rho \protect\sigma ^{3}=0.65$, $T=189.76$ $\text{K}$, and $N_{\text{eff}%
}=50000$. (D): $\protect\rho \protect\sigma ^{3}=0.5$, $T=162.93$ $\text{K}$%
, and $N_{\text{eff}}=62000$.}
\label{fig6}
\end{figure}

In Figs. \ref{fig6}.(A)-(D) we compare three different ways to calculate the
RDF. The solid line is Verlet's molecular dynamics simulation \cite{Verlet68}%
; it plays the role of experimental data against which we compare our
theory. The dotted line is $g_{hs}(r|r_{m})$ for the hard-sphere fluid with
optimal diameter $r_{m}$. This curve, calculated from eq.(\ref{G(s)}), is
also the result of the variational method and coincides with the ME result
for a macroscopically large $N_{\text{eff}}=N$. The dashed line is the
averaged $\bar{g}_{hs}(r)$ of the extended ME analysis. Figs. \ref{fig6}%
.(A)-(C) were plotted at three different temperatures $T=107.82$, $124.11%
\text{ and }189.76$ $\text{K at the density }\rho \sigma ^{3}=0.65$. Fig. %
\ref{fig6}.(D) we changed the density and the temperature to $\rho \sigma
^{3}=0.5$ and $T=162.93$ $\text{K}$. The agreement between the ME curve and
Verlet's data is good. The vast improvement over the simpler variational
method calculation is clear.

One might be tempted to dismiss this achievement as due to the adjustment of
the parameter $N_{\text{eff}}$ but this is not quite correct: $N_{\text{eff}%
} $\emph{\ has not been adjusted, it has been calculated on the basis of
information that is actually available}. Indeed, despite the fact that the
hard-sphere trial solutions that we employ are mere approximations, the
functional form of the whole curve $\bar{g}_{hs}(r)$ in eq.(\ref{ghs bar})
is reproduced quite well.

\subsection{The equation of state}

\label{sec6-4} 
\begin{figure}[ht]
\centering
\begin{picture}(0,150)(0,0)
\put(0,150){(A)}
\end{picture}
\includegraphics[width=2.0in,height=2.0in]{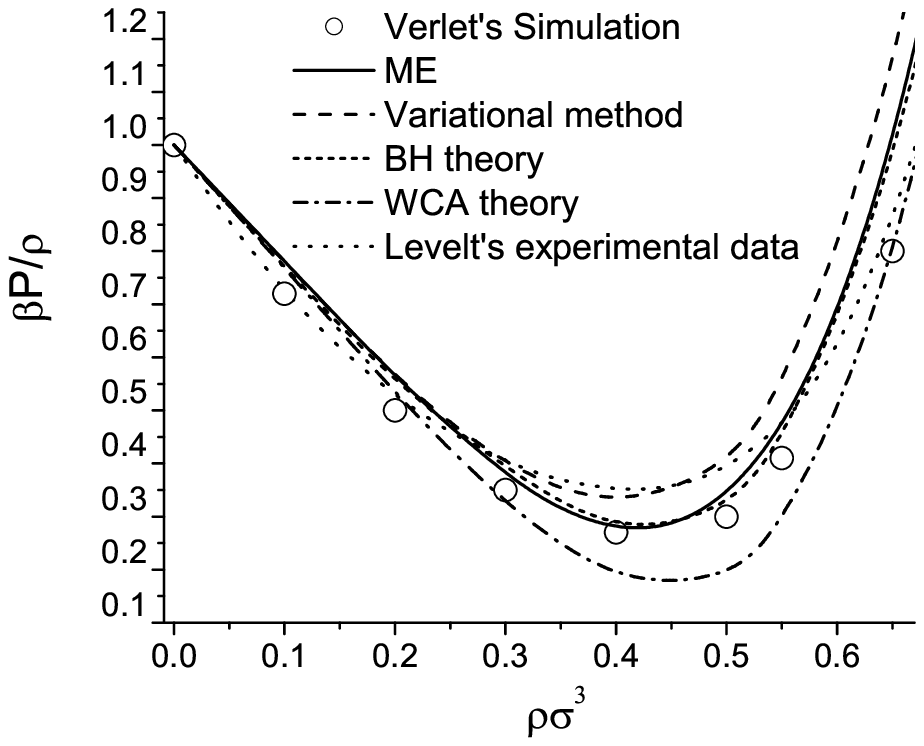} 
\begin{picture}(3,150)(0,0)
\put(3,150){(B)}
\end{picture}
\includegraphics[width=2.0in,height=2.0in]{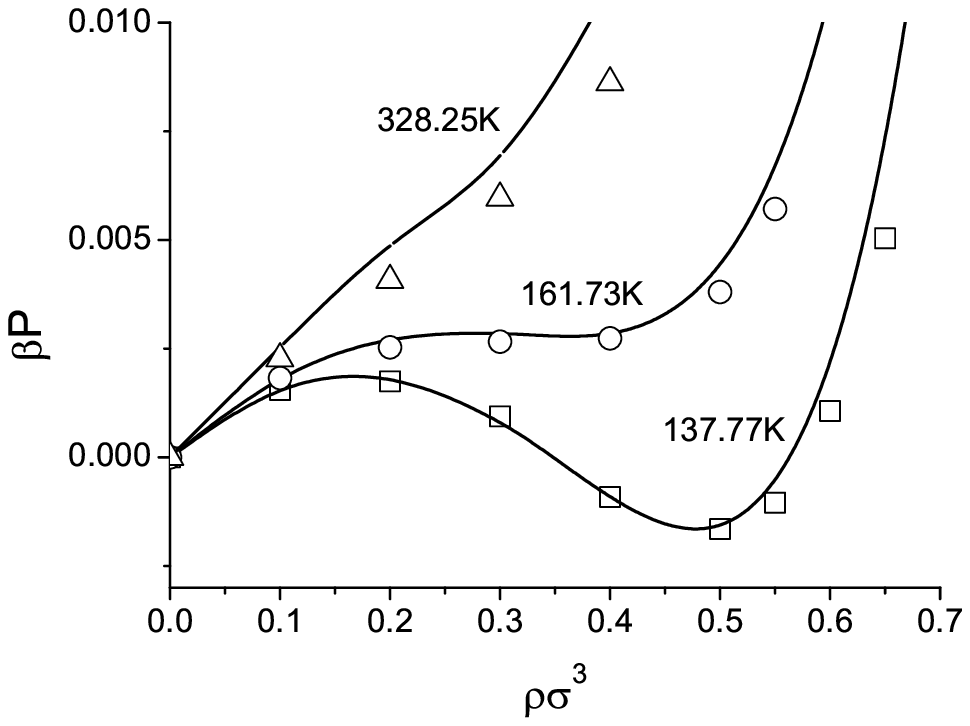}
\caption{(A): The Argon equation of state calculated using the ME method,
the variational method and the perturbative theories of BH and WCA are
compared to the Verlet simulation at $T=161.73$ $\text{K}$. Also shown are
Levelt's experimental results. (B): $\protect\beta P$ versus the reduced
density $\protect\rho \protect\sigma ^{3}$ calculated using the ME method
(solid line) and compared to the Hansen-Verlet simulation for three
different isotherms. The graph shows the appearance of the liquid-gas van
der Waals loop as the temperature drops. }
\label{fig10}
\end{figure}

Finally we use the RDF to calculate the equation of state
from the pressure equation, Eq. (\ref{eq of state}). In Fig. \ref{fig10} we
compare the equation of state derived from the $g(r)$ obtained from Verlet's
simulation with calculations using the ME and variational methods and the
perturbative theories of Barker and Henderson \cite{BarkerHenderson76} and
of Weeks, Chandler and Anderson \cite{WCA}, at $T=161.73$ $\text{K}$. The ME
results constitute a clear improvement over the plain variational
calculation. For low densities all four methods agree with each other but
differ from the simulation. A better agreement in this region would probably
require a better treatment of two-particle correlations at long distances.
At intermediate densities the best agreement is provided by the ME and BH
results, while the WCA theory seems to be the best at high densities. Also
shown in Fig. \ref{fig10}.(A) are experimental data on Argon \cite{Levelt60}%
. The discrepancy between the experimental curve and the Verlet simulation
is very likely due to the actual potential not being precisely of the
Lennard-Jones type.

In Fig. \ref{fig10}.(B) we plot the ME equation of state for three different
isotherms ($T=137.77$, $161.73$ and $328.25\text{K}$). To compare to the
simulation of Hansen and Verlet \cite{HansenVerlet69} we plot $\beta P$
(rather than $\beta P/\rho $) as a function of density $\rho \sigma ^{3}$
because this kind of plot exhibits the characteristic van der Waals loop
that signals the liquid-gas transition as the temperature drops. A more
exhaustive exploration lies, however, outside the scope of this paper.

\section{Conclusion}

\label{sec7} The goal of this paper has been to use the ME method to
generate approximations and show that this provides a generalization of the
Bogoliubov variational principle. This addresses a range of applications
that lie beyond the scope of the traditional MaxEnt. To test the method we
considered simple classical fluids.

When faced with the difficulty of dealing a system described by an
intractable Hamiltonian, the traditional approach has been to consider a
similar albeit idealized system described by a simpler more tractable
Hamiltonian. The approach we have followed here departs from this tradition:
our goal is not to identify an approximately similar Hamiltonian but rather
to identify an approximately similar probability distribution. The end
result of the ME approach is a probability distribution which is a sum or an
integral over distributions corresponding to different hard-sphere
diameters. While each term in the sum is of a form that can be associated to
a real hard-sphere gas, the sum itself is not of the form $\exp -\beta H$,
and cannot be interpreted as describing any physical system.

As far as the application to simple fluids is concerned the results achieved
in this paper represent progress but further improvements are possible by
using better approximations to the hard-sphere fluid and by choosing a
broader family of trial distributions. We argued that an important
improvement would be achieved if one could extend the trial probability
distributions to include inhomogeneities in the hard sphere diameters. This
would lead to a systematic, fully ME method for the determination of the
effective number of particles $N_{\text{eff}}$ that are locally relevant.

Many perturbative approaches to fluids had been proposed, and a gradual
process of selection over many years of research led to the optimized
theories of BH and WCA. The variational approach was definitely less
satisfactory than these ``best'' perturbation theories. With our work,
however, the situation has changed: the ME-improved variational approach
offers predictions that already are competitive with the best perturbative
theories. And, of course, the potential for further improvements of the ME
approach remains, at this early date, far from being exhausted.\newline

\begin{flushleft}
\textbf{Acknowledgements} The authors acknowledge R. Scheicher and C.-W.
Hong for their valuable assistance and advice with the numerical
calculations.
\end{flushleft}



\begin{thebibliography}{99}
\bibitem{Jaynes57} E. T. Jaynes, Information theory and statistical
mechanics, \emph{Phys. Rev.} \textbf{106}, 620-630 and Information theory
and statistical mechanics II, \emph{Phys. Rev.} \textbf{108}, 171-190
(1957); \emph{E. T. Jaynes: Papers on Probability, Statistics and
Statistical Physics}, R. D. Rosenkrantz ed. (Reidel, Dordrecht, 1983); E. T.
Jaynes, \emph{Probability Theory: The Logic of Science} (Cambridge
University Press, Cambridge, 2003).

\bibitem{ShoreJohnson80} J. E. Shore and R. W. Johnson, Axiomatic derivation
of the principle of maximum entropy and principle of minimum cross entropy, 
\emph{IEEE Trans. Inf. Theory} \textbf{IT-26}:26-36 (1980) and Properties of
cross-entropy minimization, \emph{IEEE Trans. Inf. Theory} \textbf{IT-27}%
:472-482 (1981).

\bibitem{Skilling88} J. Skilling, ``The axioms of maximum entropy'' in \emph{%
Maximum-Entropy and Bayesian Methods in Science and Engineering}, G. J.
Erickson and C. R. Smith, eds. (Kluwer, Dordrecht, 1988), pp.173-187.

\bibitem{Skilling89} J. Skilling, ``Classic maximum entropy'' in \emph{%
Maximum Entropy and Bayesian Methods}, J. Skilling, ed. (Kluwer, Dordrecht,
1989), pp.45-52.

\bibitem{Skilling90} J. Skilling, ``Quantified maximum entropy'' in \emph{%
Maximum Entropy and Bayesian Methods}, P. F. Foug\`{e}re, ed. (Kluwer,
Dordrecht, 1990), pp.341-350

\bibitem{Csiszar91} I. Csiszar, Why least square and maximum entropy?, \emph{%
Ann. Stat.} \textbf{19}:2032-2066 (1991).

\bibitem{Caticha03} A. Caticha, ``Relative entropy and inductive inference''
in \emph{Bayesian Inference and Maximum Entropy Methods in Science and
Engineering}, G. Erickson and Y. Zhai, eds. AIP Conf. Proc. \textbf{707}%
:pp.75-96 (2004) (arXiv.org/abs/physics/0311093).

\bibitem{footnote1} On terminology: The terms `prior' and `posterior' are
normally used in the context of Bayes' theorem; we retain the same
terminology when using ME because we are concerned with the similar goal of
processing information (in the form of constraints) to update from a prior
to a posterior.

\bibitem{Caticha03b} An application to general relativity is given in A.
Caticha, \textquotedblleft Towards a statistical
geometrodynamics\textquotedblright\ in \emph{Decoherence and entropy in
complex systems }, H.-T. Elze, ed.(Springer Verlag, 2004)
(arXiv.org/abs/gr-qc/0301061). An application to data analysis is given in
A. Caticha and R. Preuss, Maximum entropy and Bayesian data analysis:
Entropic prior distributions, \emph{Phys. Rev.} \textbf{E 70}%
:0461271-04612712 (2004). 

\bibitem{footnote2} Brief accounts of some of the results discussed below
have previously been presented in \cite{Tseng02} and \cite{Tseng03}.

\bibitem{Tseng02} C.-Y. Tseng and A. Caticha, ``Maximum entropy approach to
a mean field theory for fluids'' in \emph{Bayesian Inference and Maximum
Entropy Methods in Science and Engineering}, C. J. Williams, ed. AIP Conf.
Proc. \textbf{659}:pp.73-91 (2003) (arXiv.org/abs/cond-mat/0212198).

\bibitem{Tseng03} C.-Y. Tseng and A. Caticha, ``Maximum entropy approach to
the theory of simple fluids'' in \emph{Bayesian Inference and Maximum
Entropy Methods in Science and Engineering}, G. Erickson and Y. Zhai, eds.
AIP Conf. Proc. \textbf{707}:pp.17-29 (2004) (arXiv.org/abs/physics/0310746).

\bibitem{Callen85} H. B. Callen, \emph{Thermodynamics and an Introduction to
Thermostatistics} (Wiley, New York, 1985).

\bibitem{Opper01} M. Opper and O. Winther, ``From naive mean field theory to
the TAP\ equations'' in \emph{Advanced Mean Field Theory Methods}, M. Opper
and D. Sarad, eds. (MIT Press, Cambridge, 2001).

\bibitem{BarkerHenderson76} J. A. Barker and D. Henderson, ``What is
Liquid''? Understanding the states of matter, \emph{Rev. Mod. Phys.} \textbf{%
48}:587-671 (1976).

\bibitem{HansenMcDonald86} J. P. Hansen and I. R. McDonald, \emph{Theory of
Simple Liquids} (Acad. Press, 1986).

\bibitem{Kalikmanov02} V. I. Kalikmanov, \emph{Statistical Physics of Fluids 
}(Springer, 2002).

\bibitem{Mansoori69} G. A. Mansoori, F. B. Canfield, Variational approach to
the equilibrium thermodynamic properties of simple liquids I, \emph{J. Chem.
Phys.} \textbf{51}:4958-4967 (1969).

\bibitem{Stell70} G. Stell, Upper bounds on the Helmoholtz free energy, 
\emph{Chem. Phys. Lett.} \textbf{4}:651-652 (1970); J. Rasaiah and G. Stell,
Upper bounds on free energies in terms of hard-sphere results, \emph{Mol.
Phys.} \textbf{18}:249-260 (1970).

\bibitem{Lewis67} R. M. Lewis, A Unifying Principle in Statistical
Mechanics, \emph{J. Math. Phys.} \textbf{8}, 1448 (1967).

\bibitem{Karkheck82} J. Karkheck and G. Stell, Maximization of entropy,
kinetic equations, and irreversible thermodynamics, \emph{Phy. Rev. A} 
\textbf{25}:3302-3327 (1982); G. Stell, J. Karkheck, H. van Beijeren,
Kinetic mean field theories: Results of energy constraint in maximizing
entropy, \emph{J. Chem. Phys.} \textbf{79}:3166-3167 (1983); J. Karkheck, H.
van Beijeren, I. de Schepper, G. Stell, Kinetic theory and H theorem for a
dense square-well fluid, \emph{Phys. Rev. A} \textbf{32}:2517-2520 (1985);
J. Blawzdziewicz and G. Stell, Local H-theorem for a kinetic variational
theory, \emph{J. Stat. Phys.} \textbf{56}:821-840 (1989).

\bibitem{WCA} J. D. Weeks, D. Chandler and H. C. Andersen, Role of repulsive
forces in determining the equilibrium structure of simple liquids, \emph{J.
Chem. Phys.} \textbf{54}:5237-5247 (1971) and Van der Waals picture of
liquids, solids, and phase transformations, \emph{Science} \textbf{220}%
:787-794 (1983).

\bibitem{Germain02} P. Germain and S. Amokrane, Validity of the perturbation
theory for hard particle systems with very-short-range attraction, \emph{%
Phys. Rev.} \textbf{E65}:0311091-03110912 (2002).

\bibitem{Caticha00} A. Caticha, ``Maximum entropy, fluctuations and priors''
in \emph{Bayesian Inference and Maximum Entropy Methods in Science and
Engineering}, A. Mohammad-Djafari, ed. AIP Conf. Proc. \textbf{568}%
:pp.94-101 (2001) (arXiv.org/abs/math-ph/0008017).

\bibitem{Verlet68} L. Verlet, Computer ``experiments'' on classical fluids
II equilibrium corrrelation functions, \emph{Phys. Rev.} \textbf{165}%
:201-214 (1968).

\bibitem{HansenVerlet69} J.P. Hansen and L. Verlet, Phase transitions of
Lennard-Jones system, \emph{Phys. Rev.} \textbf{184}:151-161 (1969).

\bibitem{Cencov81} N. N. \v{C}encov, \emph{Statistical Decision Rules and
Optimal Inference}, Transl. Math. Monographs, vol. 53, Am. Math. Soc.
(Providence, 1981); L. L. Campbell, Proc. Am. Math. Soc. \textbf{98}:135
(1986); for applications to statistics see S. Amari, \emph{%
Differential-Geometrical Methods in Statistics} (Springer-Verlag, 1985); for
a brief derivation see A. Caticha, ``Change, time and information geometry''
in \emph{Bayesian Methods and Maximum Entropy in Science and Engineering},
A. Mohammad-Djafari, ed. AIP Conf. Proc. \textbf{568}:pp.72-85 (2001)
(arXiv.org/abs/math-ph/0008018).

\bibitem{PercusYevick58} J. K.Percus and G. J. Yevick, Analysis of classical
statistical mechanics by means of collective coordinates, \emph{Phys. Rev.} 
\textbf{110}:1-13 (1958).

\bibitem{Wertheim63} M. S. Wertheim, Exact solution of the Percus-Yevick
integral equation for hard spheres, \emph{Phys. Rev. Lett.} \textbf{10}%
:321-323 (1963) and Analytic solution of the Percus-Yevick equation, \emph{%
J. Math. Phys.} \textbf{5}:643-651 (1964); E. Thiele, Equation of state for
hard spheres, \emph{J.Chem. Phys.} \textbf{39}:474-479 (1963).

\bibitem{Bravo91} S. Bravo Yuste and A. Santos, Radial distribution function
for hard spheres, \emph{Phys. Rev. A} \textbf{43}:5418-5423 (1991); S. Bravo
Yuste, M. L\'{o}pez and A. Santos, Structure of hard-sphere metastable
fluids, \emph{Phys. Rev. E} \textbf{53}:4820-4826 (1996); Y. Tang and B.
C.-Y. Lu, Improved expressions for the radial distribution function of hard
spheres, \emph{J. Chem. Phys.} \textbf{103}:7463-7470 (1995).

\bibitem{Throop65} G. J. Throop and R. J. Bearman, Numerical solutions of
the Percus-Yevick equation for the hard-sphere potential, \emph{J. Chem.
Phys.} \textbf{42}:2408-2411 (1965).

\bibitem{Percus62} J. K. Percus, Approximation methods in classical
statistical mechanics, \emph{Phys. Rev. Lett.} \textbf{8}:462-463 (1962).

\bibitem{Levelt60} J.M.H. Levelt, The reduced equation of state, internal
energy and entropy of argon and xenon, \emph{Physica} \textbf{26}:361-377
(1960).
\end{thebibliography}
\end{document}